\documentclass[sigconf]{acmart}

\usepackage{balance}

\def\BibTeX{{\rm B\kern-.05em{\sc i\kern-.025em b}\kern-.08emT\kern-.1667em\lower.7ex\hbox{E}\kern-.125emX}}

\copyrightyear{2019}
\acmYear{2019}
\setcopyright{acmcopyright}
\acmConference[CIKM '19] {The 28th ACM International Conference on Information and Knowledge Management}{November 3--7, 2019}{Beijing, China}
\acmBooktitle{The 28th ACM International Conference on Information and Knowledge Management (CIKM'19), November 3--7, 2019, Beijing, China}
\acmPrice{15.00}
\acmDOI{10.1145/3357384.3358113}
\acmISBN{978-1-4503-6976-3/19/11}

\usepackage{amsmath}
\usepackage{amsfonts}
\usepackage{bm}
\usepackage{color}
\usepackage{graphicx}
\usepackage{url}
\usepackage{setspace}
\usepackage{subcaption}
\usepackage{booktabs}
\usepackage{mathtools}
\usepackage{multirow}
\usepackage{float}
\usepackage{caption} 
\usepackage{algorithm}
\usepackage{algorithmic}
\usepackage{cleveref}
\usepackage{makecell}
\usepackage{tabularx}
\usepackage{hyperref}

\usepackage{balance}

\crefname{equation}{Eq.}{Eq.}
\crefname{section}{Section}{Sections}
\crefname{subsection}{Section}{Sections}
\crefname{subsubsection}{Section}{Sections}
\crefname{figure}{Figure}{Figures}
\crefname{table}{Table}{Tables}
\crefname{subfigure}{Figure}{Figures}
\crefname{algocf}{Algorithm}{Algorithms}

\usepackage{enumitem}
\setlist{nosep,after=\vspace{0.0\baselineskip},leftmargin=10pt}
\setlist[itemize]{leftmargin=0.8\parindent,listparindent=\parindent,parsep=0.2\parskip,itemsep=0.02in,topsep=0.02in,after=\vspace{0in}}
\begin{document}
\fancyhead{}

\title{CosRec: 2D Convolutional Neural Networks for Sequential Recommendation}

\author{An Yan, Shuo Cheng, Wang-Cheng Kang, Mengting Wan, Julian McAuley}
\affiliation{University of California, San Diego}
\email{{ayan, scheng, wckang, m5wan, jmcauley}@ucsd.edu}

\begin{abstract}
Sequential patterns play an important role in building modern recommender systems. To this end, several recommender systems have been built on top of Markov Chains and Recurrent Models (among others).
Although these sequential models have proven successful at a range of tasks, they still struggle to uncover complex relationships nested in user purchase histories. In this paper, we argue that modeling pairwise relationships directly leads to an efficient representation of sequential features and captures complex item correlations. 
Specifically, we propose a 2D convolutional network for sequential recommendation (\textbf{CosRec}). It encodes a sequence of items into a three-way tensor; 
learns local features using 2D convolutional filters; and aggregates high-order 
interactions in a feedforward manner.

Quantitative results on two public datasets show that our method outperforms both conventional methods and recent sequence-based approaches, achieving state-of-the-art performance on various evaluation metrics. 

\end{abstract}

\maketitle

\section{Introduction}

The goal of sequential recommendation is to predict users' future behavior based on their historical action sequences. Different from traditional personalized recommendation algorithms (e.g.~Matrix Factorization \cite{koren2009matrix}) which seek to capture users' \textit{global} tastes, sequential 
models introduce 
additional behavioral dynamics by taking the \emph{order} of users' historical actions into consideration. 

\begin{figure}
	\centering
	\begin{subfigure}[b]{0.45\linewidth}
		\centering
		\includegraphics[width=\linewidth]{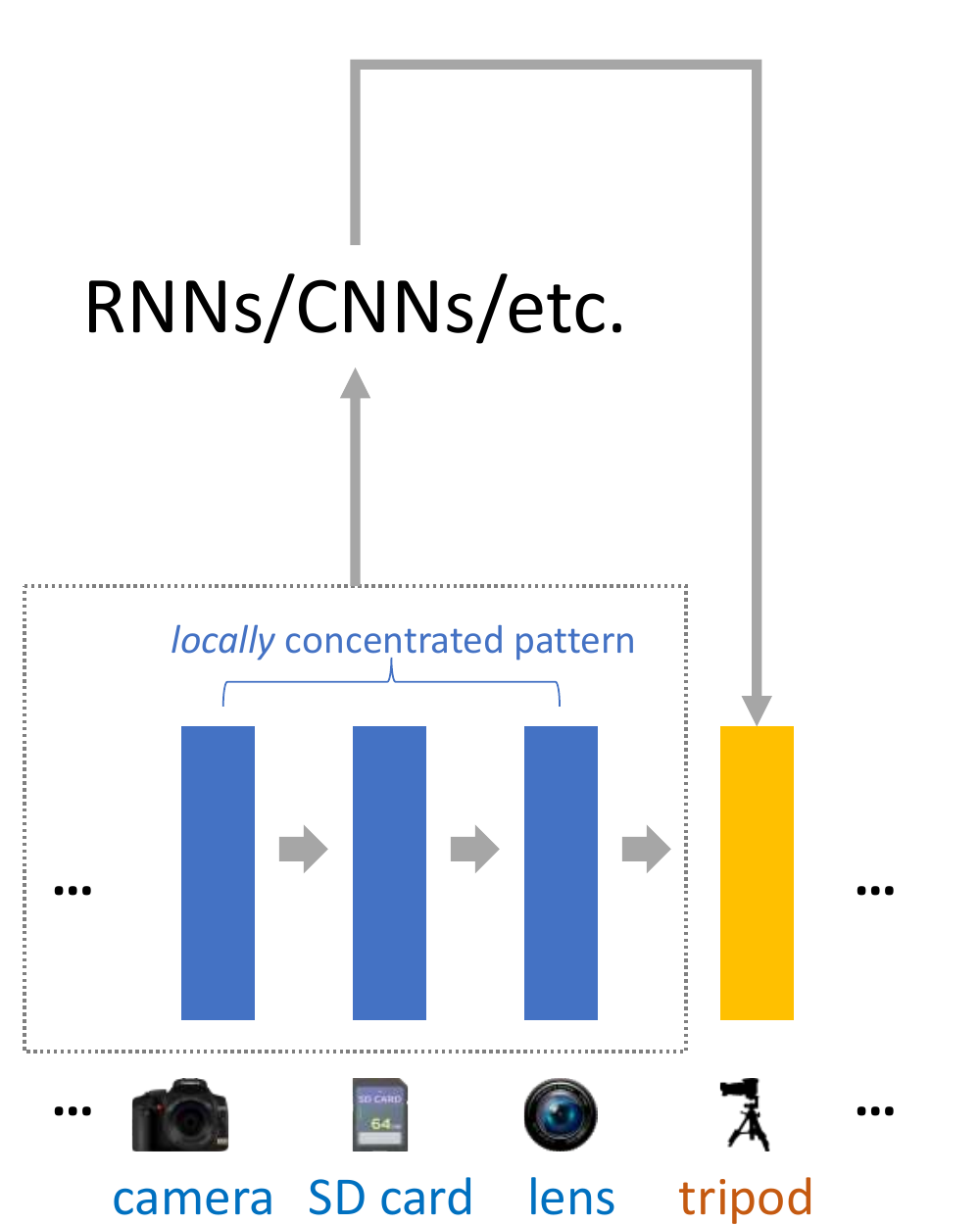}
		\vspace{-0.2in}
		\caption{}\label{fig:chain}
	\end{subfigure}
	~
	\begin{subfigure}[b]{0.45\linewidth}
		\centering
		\includegraphics[width=\linewidth]{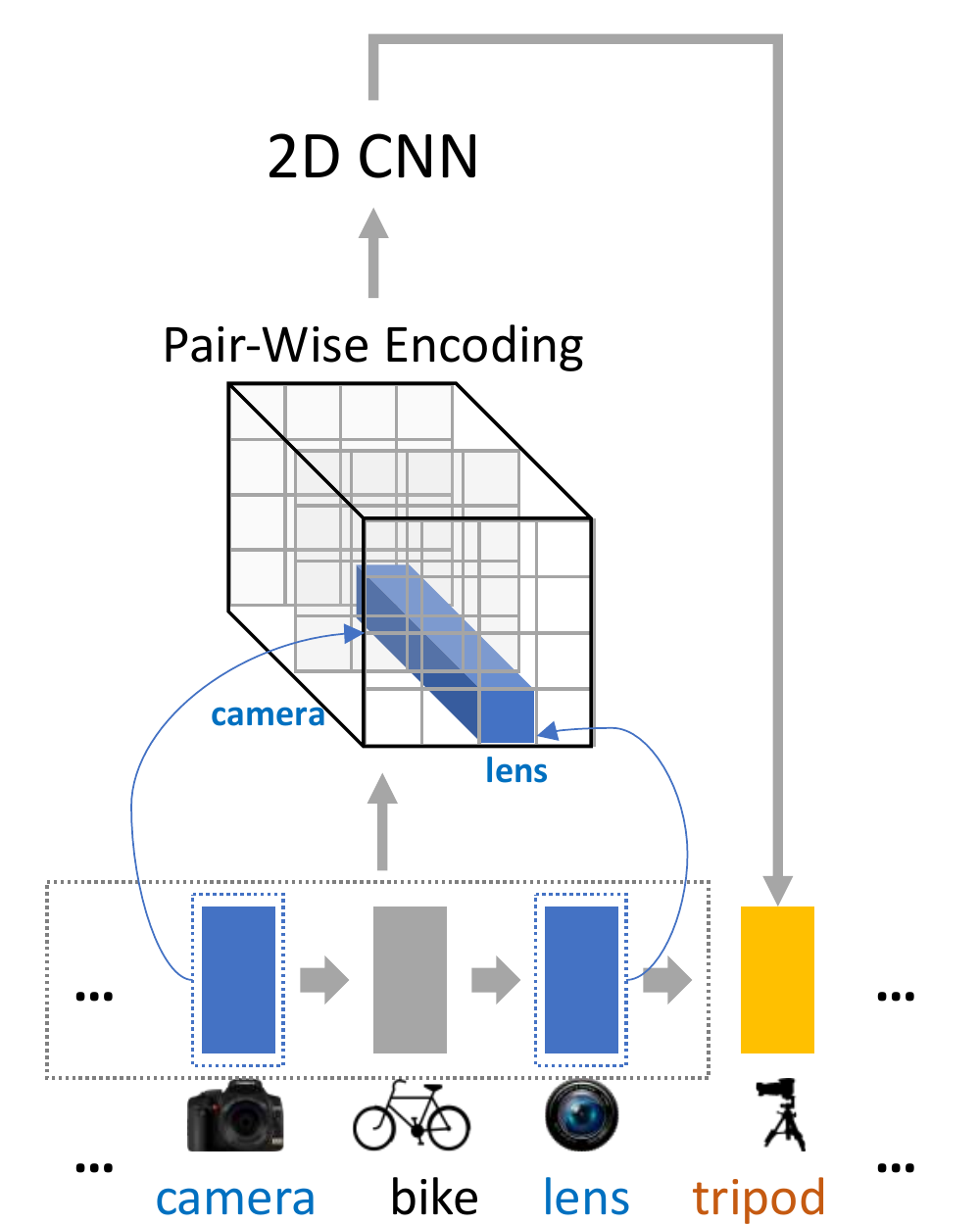}
		\vspace{-0.2in}
		\caption{}\label{fig:skip}
	\end{subfigure}
\caption{Illustrations of (a) 
locally concentrated dynamics and how they are preserved in existing models; and (b) an example where `skip' behavior (bike) exists between two closely related items (camera and lens), and how this pattern is preserved by the proposed framework.}\label{fig:compare}
\end{figure}

A classic line of work to model such dynamics is based on Markov Chains (MCs), which assumes that a user's next interaction is derived from the preceding few actions only \cite{rendle2010fpmc,he2017translation}. Recently, many neural network based approaches have achieved success on this task, where users' complete interaction sequences can be incorporated through Recurrent Neural Networks (RNNs) \cite{hidasi2015session} or Convolutional Neural Networks (CNNs) \cite{tang2018personalized}. 
Note that most existing models operate on \emph{ordered} item representations directly, and thus are constrained by the one-directional \textit{chain}-structure of action sequences. This leads to one advantage that these algorithms are capable of preserving \textit{locally} concentrated dynamics, e.g.~as shown in \cref{fig:chain}: consecutive purchases of a camera, a memory card, and a camera lens may lead to a strong indication of buying a tripod.

\textbf{In this paper}, we surprisingly find that relaxing the above structure constraint may yield more effective recommendations. Specifically, we propose a 2D CNN-based framework---2D \textbf{co}nvolutional networks for \textbf{s}equential \textbf{rec}ommendation (\textbf{CosRec}). 
In particular, we
enable interactions among \textit{nonadjacent} items by introducing a simple but effective pairwise encoding module. As shown in \cref{fig:skip}, the `skip' behavior within item sequences (i.e.,~the purchase of a bike is less relevant to the context of consuming photography products) may break the locality of the chain-structure but can be easily bypassed through this pairwise encoding. On account of this module, we show that standard 2D convolutional kernels can be applied to solve sequential recommendation problems, where small filters (e.g.~3 $\times$ 3) can be successfully incorporated. This also allows us to build an extendable 2D CNN framework, which can be easily adapted to either shallow or deep structures for different tasks. 

  \begin{figure*}[t]
  \centering
  \includegraphics[width=.75\linewidth]{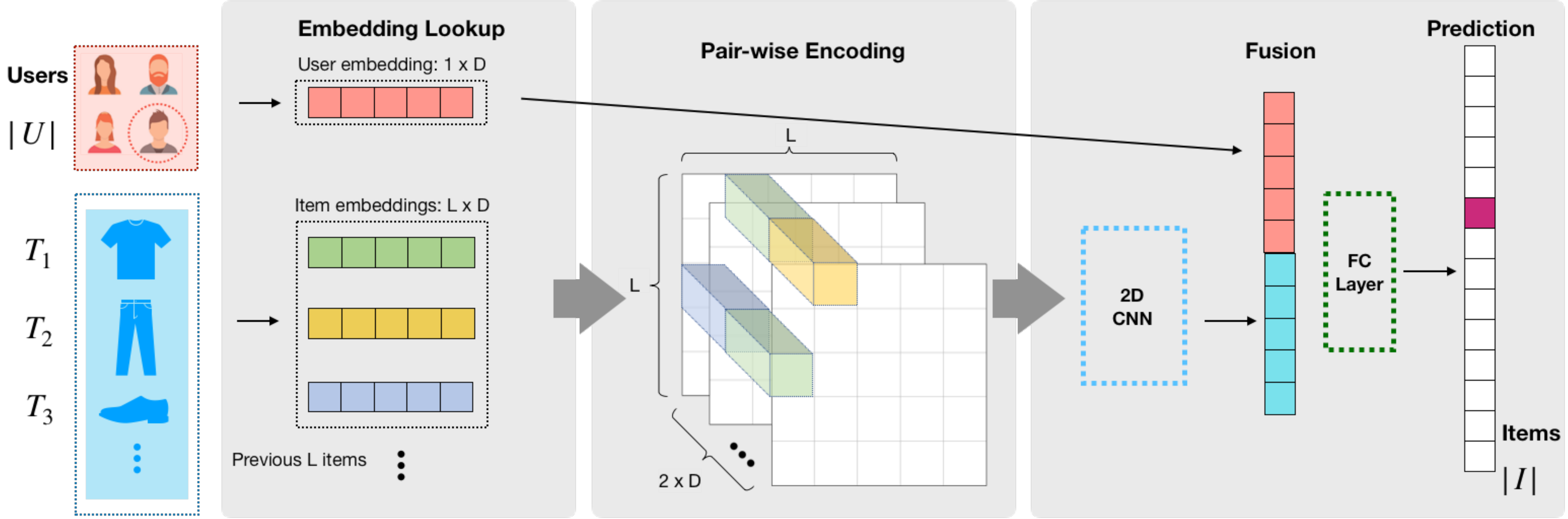}
  \caption{The detailed architecture of the proposed \textbf{CosRec} framework. Previous item embeddings (three examples illustrated here in green, yellow, and blue) are passed into a pairwise encoding module. The output is then fed into a 2D CNN and conditioned on user embeddings to predict the next item.}
  \label{fig:model}
  \end{figure*}

\section{Related Work}


Sequential recommendation methods typically seek to capture sequential patterns among previously consumed items, 
in order to accurately predict
the next item. To this end, various 
models have been adopted, including Markov Chains~(MCs)~\cite{rendle2010fpmc}, Recurrent Neural Networks~(RNNs)~\cite{hidasi2015session}, Temporal Convolutional Networks~(TCNs)~\cite{DBLP:conf/wsdm/YuanKAJ019}, and Self Attention~\cite{DBLP:conf/icdm/KangM18, sun2019bert4rec}, among others. \textbf{Caser}~\cite{tang2018personalized} and \textbf{NextItNet}~\cite{DBLP:conf/wsdm/YuanKAJ019} are closer to our work as they also use convolutions. However, \textbf{Caser}'s vertical/horizontal convolution and \textbf{NextItNet}'s 1D dilated convolution significantly differ from the standard 2D convolution used in our method, due to the different filter shapes.\footnote{\textbf{Caser}: $h\times d$, $L\times 1$, \textbf{CosRec}: $h\times h\times d$, \textbf{NextItNet}:$1\times h\times d$} To our knowledge, \textbf{CosRec} is the first 2D CNN based approach for next item recommendation.

CNNs have also been adopted for other recommendation tasks. For example, \textbf{DVBPR}~\cite{DBLP:conf/icdm/KangFWM17} extracts item embeddings from product images with CNNs, \textbf{ConvMF}~\cite{DBLP:conf/recsys/KimPOLY16} regularizes item embeddings by document features extracted from CNNs, and \textbf{ConvNCF}~\cite{DBLP:conf/ijcai/0001DWTTC18} applies CNNs on the outer product of user and item embeddings to estimate user-item interactions. Our work differs from theirs in that we focus on capturing sequential patterns of $L$ previous visited items.

\section{Method}

We formulate the sequential recommendation problem as follows. Suppose we have a set of users $\mathcal{U}$ and a set of items $\mathcal{I}$. For each user $u\in\mathcal{U}$, given the sequence of previously interacted items $\mathcal{S}^{u}=(\mathcal{S}_1^{u}, \ldots \mathcal{S}_{|S_u|}^{u})$, $\mathcal{S}_{\cdot}^u\in\mathcal{I}$, we seek to predict the next item to match the user's preferences.

We introduce our \textbf{CosRec} framework 
via
three modules: the embedding look-up layer, the pairwise encoding module, and the 2D convolution module. 
We then compare \textbf{CosRec} and other existing CNN-based approaches to illustrate how the technical limitations are addressed in our framework.
The detailed network architecture is shown in \cref{fig:model}.

\subsection{Embedding Look-up Layer}
We embed items and users into two matrices $\bm{E}_{\mathcal{I}}\in \mathbb{R}^{|\mathcal{I}| \times d}$ and $\bm{E}_{\mathcal{U}}\in \mathbb{R}^{|\mathcal{U}| \times d}$, where $d$ is the latent dimensionality, $\bm{e}_i$ and $\bm{e}_u$ denote the $i$-th and the $u$-th rows in $\bm{E}_{\mathcal{I}}, \bm{E}_{\mathcal{U}}$ respectively. Then for user $u$ at time step $t$, we retrieve the input embedding matrix $\bm{E}_{(u, t)}^{L} \in \mathbb{R}^{L \times d}$ by looking up the previous $L$ items $(\mathcal{S}_{t-L}^u,\ldots,\mathcal{S}_{t-1}^u)$ in the item embedding matrix $\bm{E}_{\mathcal{I}}$.

\subsection{Pairwise Encoding}\label{sec:pairwise}

We propose an 
encoding approach to allow flexible pairwise interactions among items. Specifically, we create a three-way tensor $\bm{T}_{(u,t)}^L \in \mathbb{R}^{L \times L \times 2d}$ on top of the input embeddings $\bm{E}_{(u,t)}^L$, where the $(i, j)$-th vector is the concatenated embedding of the item pair $(i,j)$: $[\bm{e}_i;\bm{e}_j]$, $i,j\in (\mathcal{S}_{t-L}^u,\ldots,\mathcal{S}_{t-1}^u)$.

Different from previous approaches 
\cite{tang2018personalized,DBLP:conf/wsdm/YuanKAJ019} where convolutional filters directly operate on the input matrix $\bm{E}_{(u, t)}^{L}$, we apply convolutional layers on this resulting tensor $\bm{T}_{(u,t)}^L$ so that intricate patterns (e.g.~\cref{fig:skip}) can be captured. Note that the encoded tensor has the same shape of an `image feature map' in standard CNN models for computer vision tasks. Therefore, 
a wide range of
CNN-like architectures can be borrowed and easily adapted in our context through this pairwise encoding.

\begin{table}
\setlength{\tabcolsep}{6pt}
\centering
\begin{tabular}{ccc}
\toprule
\textbf{Layer} & \textbf{Output Size} &    \textbf{Kernel Size}     \\ 
\midrule
input & $ D\times 5 \times 5$ &    -     \\ 
conv1\_1 & $D_1\times 5 \times 5$  & $1 \times 1$  \\ 
conv1\_2 & $D_1\times 3 \times 3$  & $3 \times 3$ \\ 
conv2\_1 & $D_2 \times 3 \times 3$  & $1 \times 1$  \\ 
conv2\_2 & $D_2 \times 1 \times 1$  & $3 \times 3$ \\ 
FC-$D_3$         & $D_3 \times 1 \times 1$    & dropout \\ 
FC-$|\mathcal{I}|$           & $|\mathcal{I}| \times 1 \times 1$    & sigmoid \\ 
\bottomrule
\end{tabular}
\caption{2D CNN.
$D$ is the dimension of input item embeddings. $D_1, D_2, D_3$ are the latent dimensions of each layer.
FC means Fully Connected layer.}
\label{tab:cnn}
\end{table}

\subsection{2D Convolutions}\label{sec:2dConv}
In order to capture high-level sequential patterns, we feed the above `image feature map' $\bm{T}_{(u,t)}^L$ to a 2D convolutional neural network.
We use a light-weight CNN, following suggestions from classical CNN architecture designs. 

We provide an illustrative network example in \cref{tab:cnn}. 
Here each block consists of two convolutional layers: the first layer uses $1 \times 1 $ kernels to enrich the feature representations; 
the second layer, with a kernel size of 3, 
aggregates sequential features and extracts more complex relations as the network gets deeper. Each convolutional layer is followed by a batch normalization and a rectified linear unit (ReLU) activation. 
After these two convolutional blocks, we apply a fully-connected layer with dropout and thus obtain the final sequential feature vector $\bm{v}_{(u, t)} \in \mathbb{R}^d$.

In order to capture users' global preferences, we concatenate the sequential vector $\bm{v}_{(u, t)}$ with the user embedding $\bm{e}_u$,
project them to an output layer with $|\mathcal{I}|$ nodes, and apply 
a sigmoid function 
to produce the final probability scores $\sigma(\bm{y}^{(u, t)}) \in R^{|\mathcal{I}|}$. 

\subsection{Model Training}
We adopt the binary cross-entropy loss as the objective function:
\begin{equation}
 - \sum_u \sum_t \left( log\left(\sigma(y_{\mathcal{S}^u_t}^{(u, t)})\right) + \sum_{j \notin \mathcal{S}_u} log\left(1 - \sigma(y_j^{(u, t)})\right) \right)  
\end{equation}
The network is optimized via the \textbf{Adam} Optimizer \cite{kingma2014adam}, a variant of Stochastic Gradient Descent (SGD) with adaptive moment estimation. In each iteration, we randomly sample $N$ negative samples ($j$) for each target item $\mathcal{S}^u_t$.

\subsection{Comparison with Existing CNN-based Approaches}

We show that \textbf{CosRec} addresses the limitations of existing CNN-based approaches, particularly \textbf{Caser} \cite{tang2018personalized} via the following aspects:
\begin{itemize}
    \item 
    In \textbf{Caser}, each user's action sequence is embedded as a matrix, and two types of convolutional filters are applied on top of these embeddings horizontally and vertically. One limitation of such an approach is
    that it could perform poorly in the presence of
    noisy or irrelevant interactions as shown in
    \cref{fig:skip}.
    We address this problem by encoding each action sequence into high-dimensional pairwise 
    representations and applying convolution layers afterwards, so that the above irrelevant actions can be easily skipped.
    
    \item 
    Another drawback of \textbf{Caser} is the 
    use of vertical filters. It aims to produce a weighted sum of all previous items, while it only performs summations along each dimension and there are no channel-wise interactions, which may lack representational power.
    This weighted sum also results in a shallow network structure that 
    is suited
    only for one layer, 
    leading to
    problems when modeling long-range dependencies or large-scale data streams where a deeper architecture is needed.  
    Our method with 2D kernels naturally brings channel-wise interactions among vectors, along with flexibility to adapt the network to either shallow or deep structures for different tasks, by applying padding operations or changing the kernel size.
\end{itemize}

\section{Experiments}
\subsection{Datasets and Experimental Setup}
\subsubsection{Datasets} 

Following the protocol used to evaluate \textbf{Caser} \cite{tang2018personalized}, we evaluate our method
on two standard benchmark datasets, \textbf{MovieLens} and \textbf{Gowalla}.
The statistics of
the 
datasets are shown in \cref{tab:data}.

\begin{table}
\setlength{\tabcolsep}{3pt}
\centering
\begin{tabular}{cccccc}
\toprule
\textbf{Dataset}   & \textbf{\#users} & \textbf{\#items} & \makecell[c]{\textbf{avg. \#act.}\\\textbf{per user}} & \makecell[c]{\textbf{avg. \#act.}\\\textbf{per item}}  & \textbf{\#actions}  \\
\midrule
\textbf{ML-1M}   & 6.0K     & 3.4K     & 165.50           & 292.06              & 0.993M \\ 
\textbf{Gowalla} & 13.1K    & 14.0K    & 40.74            & 38.12               & 0.533M \\ 
\bottomrule
\end{tabular}
\caption{Statistics of the datasets.}
\label{tab:data}
\end{table}

\begin{itemize}
\item{\textbf{MovieLens}} \cite{harper2016movielens}: A widely used benchmark dataset for evaluating collaborative filtering algorithms. We use the MovieLens-1M (\textbf{ML-1M}) version in our experiments.
\item{\textbf{Gowalla}} \cite{cho2011friendship}: A location-based social networking website where users share their locations by checking-in, labeled with time stamps.
\end{itemize}

We follow the same preprocessing procedure as in \textbf{Caser} \cite{tang2018personalized}:
we treat the presence of a review or rating as implicit feedback (i.e., the user interacted with the item) and use timestamps to determine the sequence order of actions, and 
discard users and items with fewer than 5 and 15 actions for \textbf{ML-1M} and \textbf{Gowalla} respectively. 
We hold the first 80\% of actions in each user's sequence for training and validation, and the remaining 20\% actions as the test set for evaluating model performance.

\subsubsection{Evaluation metrics}
We report the evaluated results by three popular top-$N$ metrics, namely Mean Average Precision (\textbf{MAP}), \textbf{Precision@N} and \textbf{Recall@N}. Here $N$ is set to 1, 5, and 10. 
\begin{table*}[t]
\centering
\begin{tabular}{ccccccccccc}
\toprule
\textbf{Dataset} 				& \textbf{Metric }  & \textbf{PopRec} & \textbf{BPR}   & \textbf{FMC}    & \textbf{FPMC}   & \textbf{GRU4Rec} & \textbf{Caser}  & \textbf{CosRec-base}   & \textbf{CosRec} & \textbf{Improvement} \\ 
\midrule
\multirow{7}{*}{$ML-1M$}  
& MAP      &0.0687 &0.0913 &0.0949 &0.1053 &0.1440 &0.1507 &0.1743 &\textbf{0.1883} & +25.0\%\\
& Prec@1   &0.1280 &0.1478 &0.1748 &0.2022 &0.2515 &0.2502 &0.2892 &\textbf{0.3308} & +31.5\%\\ 
& Prec@5   &0.1113 &0.1288 &0.1505 &0.1659 &0.2146 &0.2175 &0.2521 &\textbf{0.2831} & +30.2\%\\
& Prec@10  &0.1011 &0.1193 &0.1317 &0.1460 &0.1916 &0.1991 &0.2256 &\textbf{0.2493} & +25.2\%\\
& Recall@1 &0.0050 &0.0070 &0.0104 &0.0118 &0.0153 &0.0148 &0.0186 &\textbf{0.0202} & +32.0\%\\
& Recall@5 &0.0213 &0.0312 &0.0432 &0.0468 &0.0629 &0.0632 &0.0771 &\textbf{0.0843} & +33.4\%\\
& Recall@10&0.0375 &0.0560 &0.0722 &0.0777 &0.1093 &0.1121 &0.1331 &\textbf{0.1438} & +28.3\%\\ 
\midrule
\multirow{7}{*}{$Gowalla$}  
& MAP      &0.0229 &0.0767 &0.0711 &0.0764 &0.0580 &0.0928 &0.0821 &\textbf{0.0980} & +05.6\%\\
& Prec@1   &0.0517 &0.1640 &0.1532 &0.1555 &0.1050 &0.1961 &0.1712 &\textbf{0.2135} & +08.9\%\\ 
& Prec@5   &0.0362 &0.0983 &0.0876 &0.0936 &0.0721 &0.1129 &0.1012 &\textbf{0.1190} & +05.4\%\\ 
& Prec@10  &0.0281 &0.0726 &0.0657 &0.0698 &0.0782 &0.0571 &0.0762 &\textbf{0.0884} & +13.0\%\\
& Recall@1 &0.0064 &0.0250 &0.0234 &0.0256 &0.0155 &0.0310 &0.0265 &\textbf{0.0337} & +08.7\%\\
& Recall@5 &0.0257 &0.0743 &0.0648 &0.0722 &0.0529 &0.0845 &0.0752 &\textbf{0.0890} & +05.3\%\\
& Recall@10&0.0402 &0.1077 &0.0950 &0.1059 &0.0826 &0.1223 &0.1107 &\textbf{0.1305} & +06.7\%\\ 
\bottomrule
\end{tabular}
\caption{Performance comparison with state-of-the-art approaches on all datasets.}
\label{tab:perform}
\end{table*}

\subsubsection{Implementation details} 
We use 2 convolution blocks, each consisting of 2 layers.
The latent dimension $d$ is chosen from \{10, 20, 30, 50, 100\}, and we use 50 and 100 for \textbf{ML-1M} and \textbf{Gowalla} respectively. The Markov order $L$ is 5. We predict the next $T=3$ items at once. The learning rate is 0.001, with a batch size of 512, a negative sampling rate of 3 and a dropout rate of 0.5. All experiments are implemented using PyTorch.%
\footnote{Code is available at: \url{https://github.com/zzxslp/CosRec}}

\subsection{Performance Comparison}
To show the effectiveness of our method, we compare it with a number of popular baselines.
\begin{itemize}
\item{\textbf{PopRec}:} A simple baseline that ranks items according to their popularity. 

\item{\textbf{Bayesian Personalized Ranking (BPR)} \cite{rendle2009bpr}:} A classic non-sequential method for learning personalized rankings.
\item{\textbf{Factorized Markov Chains (FMC)} \cite{rendle2010fpmc}:} A first-order Markov Chain method which generates recommendations depending only on the last visited item.

\item{\textbf{Factorized Personalized Markov Chains (FPMC) \cite{rendle2010fpmc}:}} A combination of FMC and MF,
so that short-term item transition patterns as well as users' global preferences can be captured. 

\item{\textbf{GRU4Rec} \cite{hidasi2015session}:} A state-of-the-art model which uses an RNN to capture sequential dependencies and make predictions.

\item{\textbf{Convolutional Sequence Embeddings (Caser)} \cite{tang2018personalized}:} A recently proposed CNN-based method which captures high-order Markov Chains by applying convolutional operations on the embedding matrix of the previous $L$ items.

\item{\textbf{CosRec-base}:}
In order to evaluate the effectiveness of the 2D CNN module, we create
a baseline version of \textbf{CosRec} 
which uses a multilayer perceptron (MLP) instead of a 2D CNN on the pairwise encodings. 
\end{itemize}
Experimental results are summarized in \cref{tab:perform}.
Among the baselines, sequential models (e.g.~\textbf{Caser}) outperform non-sequential models (e.g.~\textbf{BPR}), confirming the importance of considering sequential information. \textbf{CosRec} outperforms \textbf{FMC}/\textbf{FPMC}, since they only model the first-order Markov chain while \textbf{CosRec} captures 
high-order relations.
Overall, our method outperforms all baselines on both datasets 
by a significant margin. 
The performance improvements
on \textbf{ML-1M} are particularly 
significant (\textbf{25.0\%} improvement in terms of \textbf{MAP}), 
presumably due to the fact that
\textbf{ML-1M} is a relatively dense 
dataset with rich 
sequential signals. 
Note on the \textbf{ML-1M} dataset, even our baseline version (MLP) 
outperforms existing state-of-the-art methods, which validates the effectiveness of the pairwise encoding for capturing more intricate patterns. 

\begin{figure}
	\centering
	\begin{subfigure}[b]{0.475\linewidth}
		\centering
		\includegraphics[width=\linewidth]{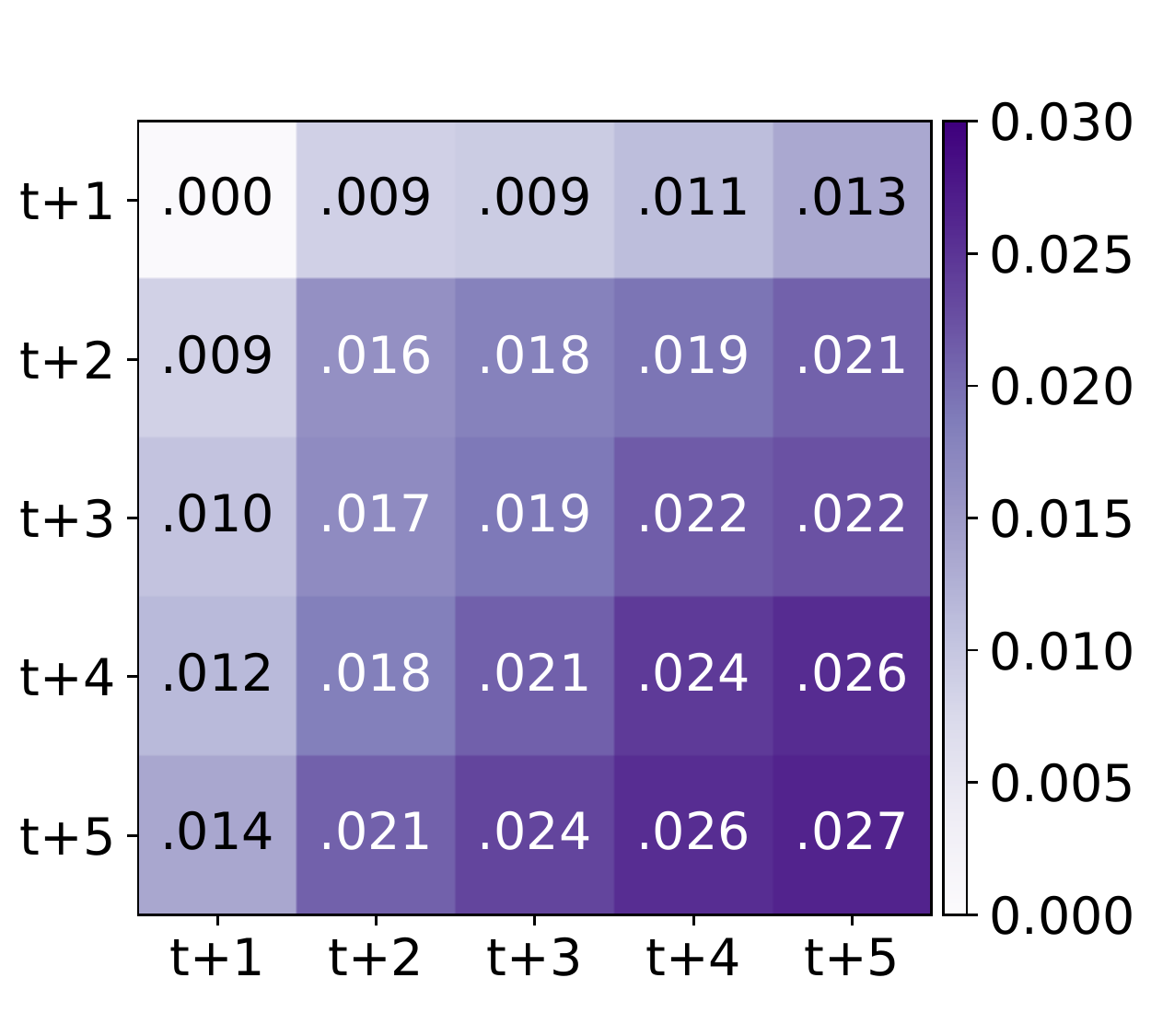}
		\caption{}\label{fig:left}
	\end{subfigure}
	~
	\begin{subfigure}[b]{0.475\linewidth} 
		\centering
		\includegraphics[width=\linewidth]{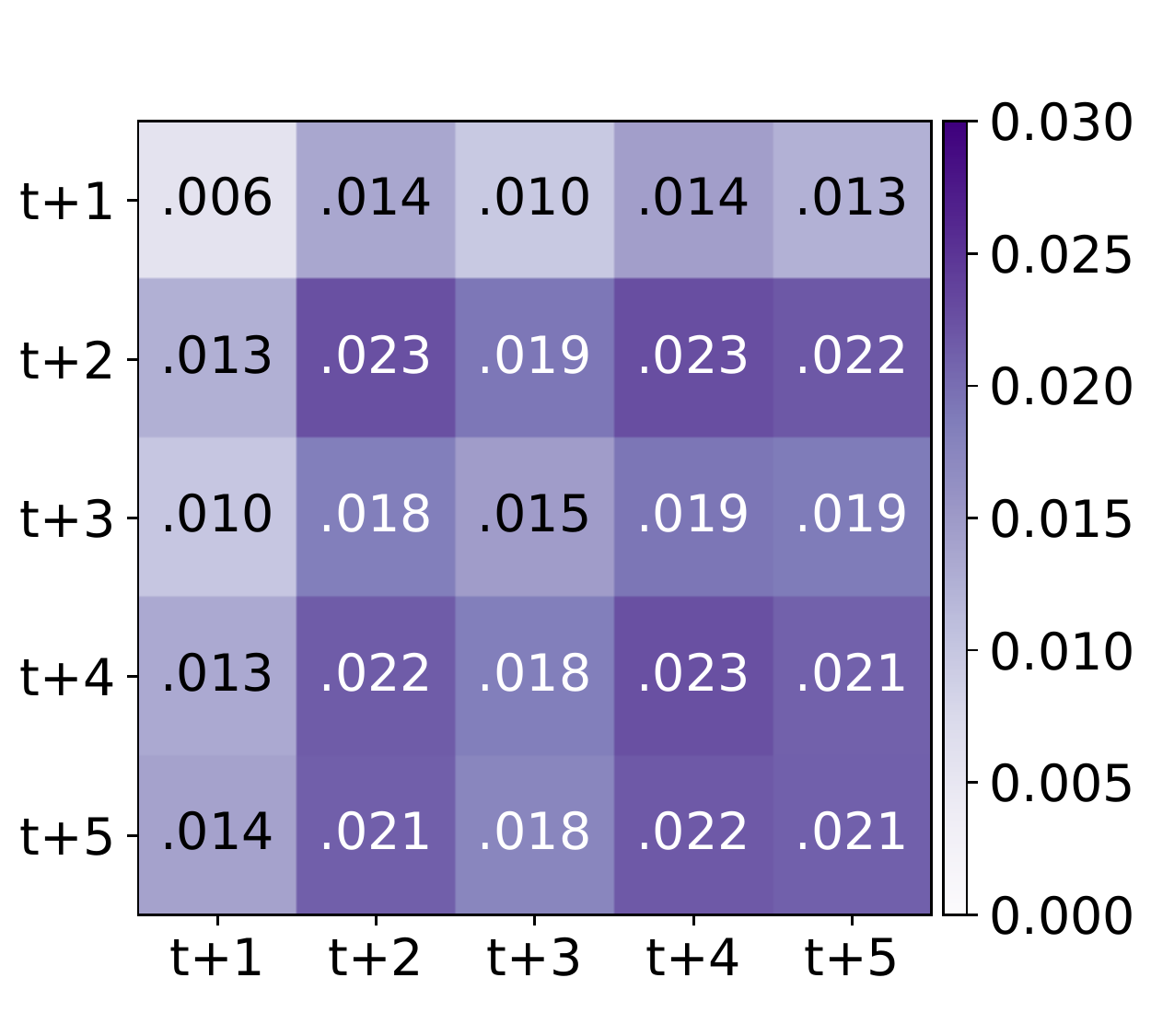}
		\caption{}\label{fig:right}
	\end{subfigure}
\caption{A visualization of two convolutional filters with kernel size 5, trained on ML-1M. Darker colors indicate higher values.
The value in grid (i,j) corresponds to the weight for the item pair (i,j)'s pairwise encoding.
}
\label{fig:visual}
\end{figure}

\subsection{Visualization}
We visualize two example convolutional filters in \cref{fig:visual} to show that our proposed framework is not only capable of modeling the `recency' and the `locality' (as in existing models), but flexible to capture more complex patterns such as the `skip' behavior. Here each filter serves as a weighted sum of our pairwise encoding. We see a clear trend in \cref{fig:left} that weights increase from top left to bottom right, which indicates more recent items are attended in this case. In addition, we observe scattered blocks in \cref{fig:right}, which implies that the model is able to bypass the chain-structure and capture nonadjacent patterns.

\section{Conclusion}
In this paper, we proposed a novel 2D CNN framework, \textbf{CosRec} for 
sequential recommendation. 
The model encodes a sequence of item embeddings into pairwise representations and leverages a 2D CNN to extract 
sequential features. 
We perform 
experiments on two real-world datasets, where our method significantly outperforms recent sequential approaches, showing its effectiveness as a generic model for sequential recommendation tasks.

\noindent\textbf{Acknowledgements.} This work is partly supported by NSF \#1750063. We thank all the reviewers for their 
suggestions.

\bibliographystyle{ACM-Reference-Format}
\balance
\bibliography{cikm19}

\end{document}